\documentstyle[prb,aps,psfig]{revtex}
\begin{document}
\draft
\title{Scaling properties of the critical behavior in the
dilute antiferromagnet $\mathbf Fe_{0.93}Zn_{0.07}F_2$}
\author{Z. Slani\v{c} and D. P. Belanger}
\address{Department of Physics, University of California,
Santa Cruz, CA 95064 USA}
\author{J. A. Fernandez-Baca}
\address{Solid State Division, Oak Ridge National Laboratory,
Oak Ridge, TN 37831-6393 USA}
\date{\today}
\maketitle
\begin{abstract}
Critical scattering analyses for dilute antiferromagnets
are made difficult by the lack of predicted theoretical line
shapes beyond mean-field models.  Nevertheless, with the use
of some general scaling assumptions we have
developed a procedure by which we can analyze
the equilibrium critical scattering in these systems
for $H=0$, the random-exchange Ising model, and, more importantly,
for $H>0$, the random-field Ising model.
Our new fitting approach, as opposed to the more conventional
techniques, allows us to obtain the universal critical behavior
exponents and amplitude ratios as well as the critical line shapes.
We discuss the technique as applied to $Fe_{0.93}Zn_{0.07}F_2$.
The general technique, however, should be
applicable to other problems where the scattering line shapes
are not well understood but scaling is expected to hold.
\end{abstract}

\section{Introduction}

Characterizations of the critical behavior of
model systems through experiments and simulations
are essential to verify the validity of theoretical
models of phase transitions.
Scattering techniques are invaluable for characterizing
the staggered magnetization (the order parameter), $M _s$,
the antiferromagnetic fluctuation correlation length, $\xi$, and 
staggered susceptibility, $\chi _s^{}$, as a function of
temperature in pure and dilute antiferromagnets, which
prove to be ideal physical realizations of many model
systems.  Neutron scattering has been particularly
instrumental in studies of antiferromagnets\cite{c89},
though magnetic x-ray scattering has also
been employed to a limited extent\cite{gmshgptb87}.  Likewise,
pulsed heat and optical techniques\cite{fg84} have
been essential in determining the critical
behavior of the specific heat, $C_m$.
The Ising
model is the simplest of systems, with each
spin having only two possible states, and becomes
an exact model for the anisotropic antiferromagnets
as the temperature $T$ approaches the transition temperature
$T_c$.
Three of the most fundamental phase transitions
are the pure Ising model, the random-exchange
Ising model (REIM) and the random-field Ising model (RFIM).
The pure Ising model has the exact Onsager\cite{o44} solution
for dimension $d=2$.  For $d=3$ only approximate renormalization
solutions, simulations and expansions
exist.  Antiferromagnets exist with
strong anisotropy, such as $FeF_2$ for $d=3$.
This system exhibits
universal Ising static critical behavior
close to the transition temperature\cite{bnkjlb83,by87,wb67}.  The
random-exchange model system is
realized when the magnetic ions are randomly
substituted by diamagnetic ions in
$Fe_xZn_{1-x}F_2$ for $x>x_p$, where $x_p = 0.246$ is the percolation
threshold for this body-centered tetragonal magnetic lattice
if only the dominant next-nearest neighbor interaction\cite{lz98}
is considered.
Below $x_p$ no phase transition can occur for geometric connectivity
reasons.  The anisotropy increases as $x$ decreases\cite{a80}.
As a result of the high anisotropy for these systems measured
using neutron scattering\cite{hrg70},
asymptotic Ising critical behavior is well followed in the
reduced temperature range $|t| = |(T-T_c)/T_c| < 10^{-2}$, where $T_c$ is
the transition temperature\cite{by87}.

Various experimental techniques can be used to
extract universal Ising parameters associated
with the asymptotic critical behaviors in antiferromagnets\cite{by91}.
The universal parameters accessible through scattering techniques
include the exponents and amplitude ratios
associated with the asymptotic power law behaviors
\begin{equation}
\xi = \xi _o^\pm |t|^{-\nu} \quad ,
\end{equation}
where $+$ and $-$ refer to $t>0$ and $t<0$, respectively,
\begin{equation}
M _s = M _o|t|^{\beta} \quad,
\end{equation}
where $M_o$ is nonzero for $t<0$ only,
\begin{equation}
\chi _s^{} = \chi _o^\pm |t|^{-\gamma} \quad ,
\end{equation}
and, for antiferromagnets with quenched randomness,
the disconnected susceptibility
\begin{equation}
\chi _s^{dis} = \chi _o^{dis \pm} |t|^{-\bar\gamma}
\end{equation}
for $q \ne 0$.  Pulsed heat and optical techniques can be used
to obtain the specific heat critical behavior,
\begin{equation}
C = A^{\pm}|t|^{-\alpha} +B \quad ,
\end{equation}
which becomes symmetric and logarithmic,
\begin{equation}
C = A\ln |t| \quad ,
\end{equation}
when $\alpha \rightarrow 0$.

Equations 1-6 represent the asymptotic behaviors exhibited for
data with sufficiently small $|t|$.  The range of asymptotic Ising
behavior in anisotropic short-range interaction antiferromagnets,
for example, is partly determined by the anisotropy strength.
For example, whereas $FeF_2$ shows\cite{bnkjlb83} asymptotic
behavior for $|t|<10^{-2}$, the
less anisotropic isomorph $MnF_2$\cite{bjr78,nld69} shows
asymptotic behavior only
for $|t|<10^{-3}$.  If data are taken outside the asymptotic
range of $|t|$, but still within the critical region,
fits to the power law expressions yield only
effective exponents\cite{aa80} and effective amplitude ratios.  In such cases,
it may be more effective to use scaling function analyses, which
include crossover to asymptotic critical behaviors, when some
quantity such as the applied field, $H$, can be varied.  We will
briefly discuss this with regard to specific heat analyses
done principally using $Fe_{0.93}Zn_{0.07}F_2$ data.
Scaling functions are also of great utility in fitting scattering
data when the line shapes are not well known.  For example,
at constant $H$ in the asymptotic region, we can use the fact
that the scaling functions can depend only on $|q|/\kappa$, where
$q$ is the distance from the antiferromagnetic Bragg scattering point,
$(1 0 0)$ for $FeF_2$, in reciprocal lattice units (rlu) and
$\kappa = 1/\xi$ is the reciprocal correlation length for
antiferromagnetic fluctuations.  In this work,
we will show how the general properties of the scaling functions
can be utilized to advantage in the characterization of the random-exchange
and random-field scattering data obtained\cite{sbf99}
using $Fe_{0.93}Zn_{0.07}F_2$.
However, the technique has more general utility; it can be used in
any case where theory does not provide adequate models of scattering
functions but scaling is expected to hold.

\section{Specific heat scaling}

Although we primarily focus here on the scaling behavior
of the critical scattering of the $d=3$ dilute
antiferromagnet, it is instructive to review
briefly the success of scaling analyses in the study of the
specific heat critical behavior in the same system.
Not only will this demonstrate the usefulness of the
scaling approach, it will highlight the results
for the specific heat behavior, which are complementary to
the scattering results but show the largest discrepancy with simulation
results for the RFIM.  This will be important in our later assessment of the 
agreement between simulations and experimental results.

In zero field, the dilute anisotropic antiferromagnet is
predicted to have a transition described by the
random-exchange Ising model.  One of the most striking
changes in the critical behavior induced by random
quenched dilution is observed in the specific heat for $d=3$.  Whereas
the pure $FeF_2$ sample shows $\alpha=0.11$, in agreement
with theory\cite{cprv99}, the Harris criterion imposes the constraint
that $\alpha < 0$ upon dilution.  Indeed, the exponent has been found\cite{sb98}
to be $\alpha = -0.10 \pm 0.02$ experimentally.  Monte Carlo
studies also yield a negative value\cite{bfmspr98}.  Interestingly,
with the application of a field,
the RFIM specific heat is again found to be divergent
in experiments\cite{sb98}, with $\alpha \approx 0$,
as discussed below.

One way to utilize scaling functions is to 
attempt to collapse the experimental data onto a scaling function
of the appropriate scaling argument.  The collapse will only
work well if the critical parameters used in the data collapse are correct.
The RFIM scaling behavior of the free energy, for example, is expected
to have the form\cite{by91}
\begin{equation}
F \sim H^{2(2- \alpha)/\phi _{RF}}g(|t_H|H^{-2/\phi_{RF}}) \quad ,
\end{equation}
where $t_H = (T-T_N+bH^2)/T_N$, $T_N$ is the zero-field transition temperature,
$b$ is a small mean-field (MF) parameter, $\alpha$ is the zero-field
specific heat exponent, and $\phi_{RF}$ is the RFIM crossover exponent.
When a new phase transition occurs at $T=T_c(H)$, the asymptotic limit
of the specific heat can be obtained\cite{kkj86} from the free energy in
the limit $|t| \rightarrow 0$
\begin{equation}
C = \frac {\partial ^2 F}{\partial T ^2} = H^{-2\alpha /\phi _{RF}}g'(|t_H|H^{-2/\phi_{RF}}) \sim H^{2(\bar\alpha-\alpha) /\phi_{RF}} |t|^{-\bar\alpha} \quad ,
\end{equation}
where $t=(T-T_c(H))/T_c(H)$.
For $\bar\alpha = 0$, this becomes
\begin{equation}
C  \sim H^{-2\alpha /\phi_{RF}} \ln |t| \quad , 
\end{equation}
which is symmetric above and below $T_c(H)$.  Note that the field
dependence of the peak amplitude is dependent on $\bar\alpha$.
In the scaling plots, both the shape and the field
dependent amplitude must be correct for the data to collapse onto
a single scaling function.

The scaling behavior of the specific heat has been experimentally
demonstrated for both $d=2$, where no transition takes
place\cite{fkjcg83},
and $d=3$, where a new transition occurs\cite{sb98}.
In both cases, the critical
parameters can be determined accurately by the
quality of the data collapse onto a single scaling function.
The most accurate measurements have been obtained using the
optical linear birefringence technique which has been
shown to faithfully represent the magnetic specific heat
behavior\cite{fg84,sb98,db89}.
It was shown that the experimental data for the $d=3$
system $Fe_{0.93}Zn_{0.07}F_2$, when
divided by $H^{-2\alpha/\phi_{RF}}$, collapse onto a single
scaling function if $\alpha = -0.10 \pm 0.02$ and
$\phi_{RF} = 1.42 \pm 0.03$, the latter having been determined by measurements
on $Fe_{x}Zn_{1-x}F_2$ with several different
concentrations\cite{fkj91} and
predicted\cite{a86} to be a few percent larger than the zero-field staggered
susceptibility exponent measured\cite{bkj86}
to be $\gamma = 1.31 \pm 0.03$.
Note that if the RFIM  specific heat exponent $\bar\alpha = 0$,
as indicated by 
the asymptotic shape of the curve, the random-exchange exponent must
have the value $\alpha = -0.10 \pm 0.02$, which is consistent with the
earlier value $\alpha = -0.09 \pm 0.03$ obtained\cite{bcsybkj83} by fitting
the data from a lower magnetic concentration sample with $x=0.60$ to a
power law.   Hence, both the shape and
field-dependent amplitude of the critical peak are consistent with
$\bar\alpha = 0$, i.e.\ a symmetric, logarithmic divergence.

Further evidence that $\bar\alpha$ is close to zero is
obtained from Faraday rotation experiments 
under constant field or constant temperature.  Faraday measurements
yield the critical behavior of the specific heat, but with a
different field amplitude than the specific heat peaks.
Near the phase transition it was shown
by Kleemann, et al. \cite{kkj86}, that
\begin{equation}
\frac {\partial M}{\partial T} = \frac{\partial ^2 F}{\partial H \partial
T} \sim H^{2(1+\bar\alpha - \alpha -\phi
_{RF}/2)/\phi _{RF}} |t| ^{-\bar\alpha} \quad ,
\end{equation}
which becomes, for $\bar\alpha = 0 $,
\begin{equation}
\frac {\partial M}{\partial T} \sim H^{2(1 - \alpha -\phi
_{RF}/2)/\phi _{RF}} \ln |t| \quad ,
\end{equation}
and
\begin{equation}
\frac {\partial M}{\partial H} = \frac {\partial ^2 F}{\partial H ^2} \sim H^{2(2+\bar\alpha - \alpha -\phi _{RF})/\phi _{RF}} |t| ^{-\bar\alpha} \quad ,
\end{equation}
which becomes, for $\bar\alpha = 0 $,
\begin{equation}
\frac {\partial M}{\partial H} \sim H^{2(2 - \alpha -\phi _{RF})/\phi _{RF}} \ln |t| \quad .
\end{equation}
The field dependent amplitudes of logarithmic peaks have been measured
with the results $2(1 - \alpha -\phi _{RF}/2)/\phi _{RF} \approx 0.56$
and $2(2 - \alpha -\phi _{RF})/\phi _{RF} \approx 0.97$.
With the measured value $\phi _{RF} = 1.42 \pm 0.03$, these
two equations yield $\alpha = -0.11 \pm 0.02$ and $\alpha = -0.11 \pm 0.04$,
respectively.  These values are consistent with the values
$\alpha = -0.10 \pm 0.02$ from the $Fe_{0.93}Zn_{0.07}F_2$
experiment\cite{sb98}
and $\alpha = -0.09 \pm 0.03$ from the $Fe_{0.60}Zn_{0.40}F_2$
experiment\cite{bcsybkj83}.  Hence, the field dependent amplitudes as well
as the peak shapes are all consistent with $\bar\alpha \approx 0$.
A scaling analysis would similarly require $\bar \alpha \approx 0$ in the
case of Faraday rotation for a good data collapse.
The result $\bar \alpha \approx 0$ from experiments contradicts the
Monte Carlo result\cite{r95} $\bar \alpha = -0.5 \pm 0.2$.

\section{Scattering scaling function}

We next turn to the scattering function, which should
obey scaling properties close to $T_c(H)$.
Within the static approximation\cite{c89},
the intensity of the magnetic critical scattering
from high quality single crystal
magnetic systems is proportional to the Fourier
transform of the spin-spin correlation function $S(q)=[<s_qs_{-q}>]$
convoluted with the instrumental resolution, where the $<>$ brackets
signify a thermal average and the $[$ $]$ brackets signify a configurational
average.

The spin-spin correlation can be expressed as
\begin{equation}
S(q)= \chi _s^{} + \chi _s^{dis} \quad ,
\end{equation}
where 
\begin{equation}
\chi _s^{}= [<s_qs_{-q}>-<s_q><s_{-q}>]
\end{equation}
is the staggered susceptibility and
\begin{equation}
\chi _s^{dis} = [<s_q><s_{-q}>]
\end{equation}
is the disconnected susceptibility.  For line shapes obtained at one
value of the field, we expect
a scaling function that only depends on the ratio $|q|/\kappa$
of the two physically relevant inverse length scales.  For $|q|>0$,
\begin{equation}
{\chi _s^{}}(q)=A^{\pm}\kappa ^{\eta -2}f(q/\kappa) \quad .
\end{equation}
For pure, translationally
invariant systems $\chi _s^{dis}={M_s}^2\delta (q)$.  For random systems,
on the other hand, $\chi _s^{dis}$ may have a $q$ dependent contribution.
In particular, in RFIM systems, such a term is induced by the random
field, giving\cite{by91} for $|q|>0$ at constant $H$,
\begin{equation}
\chi _s^{dis}(q)=A^{\pm}B^{\pm}\kappa ^{\bar{\eta} -4}g(q/\kappa) \quad .
\end{equation}
Hence, $S(q)$ involves two possibly independent scaling functions,
$f(q/\kappa)$ and $g(q/\kappa)$ and two possibly independent exponents
$\gamma$ and $\bar\gamma$.  This makes the method of
collapsing data onto scaling functions extremely difficult since
the contributions to the data from the two scaling functions
are not easily separable.

To fit the data we must, in principle, use data only in the
range of small $\kappa$ (i.e. small $|t|$) and $|q|$ in order to
be sure we are dealing with asymptotic behavior.  In the study
of $FeF_2$, data were used in the range $|t|<10^{-2}$ to obtain
the most reliable exponents and amplitude ratios, since
this was shown to be in the asymptotic range for pure
Ising behavior in specific
heat critical behavior\cite{bnkjlb83} measurements.  In the
$Fe_{0.93}Zn_{0.07}F_2$ experiments, we used data for $|t|<10^{-2}$
since this is the range for which the specific heat shows the
RFIM logarithmic behavior\cite{sb98}.
However, we typically use a wide range of $|q|$. The crossover
at large $|q|$ is relatively unimportant since
the critical scattering intensity becomes very small.  On the other hand,
including data at large $|q|$ helps to set the level of background scattering.

Many scattering critical behavior analyses are done using
the simple MF Lorentzian for $|q|>0$,
\begin{equation}
f(q)=\frac {A^\pm}{1+q^2/\kappa ^2} \quad ,
\end{equation}
giving
\begin{equation}
S(q)=\chi _s^{} (q) = \frac {A^{\pm}}{q^2+\kappa ^2}
\end{equation}
for $q \ne 0$, consistent with the fact that $\eta = 0 $ in MF.
Since the upper critical dimension, $d_u$, above which the
MF equations are correct, is four or greater,
the Lorentzian line shape can only be approximate for
three dimensions.  Deviations from the MF Lorentzian
line shape should become evident as one approaches
the transition temperature and are generally found to be much more important
below the transition temperature.  This has been
discussed with respect to the pure $d=3$ Ising antiferromagnet
$FeF_2$\cite{by87}.  Scattering data for $FeF_2$ were analyzed
using approximations to the line shapes by Fisher and
Burford\cite{fb67} (BF) for $T>T_N$,
\begin{equation}
f(q/\kappa ) \propto \frac
{(1+\phi^{2}q^{2}/\kappa ^{2})^{\eta/2}}
{1+\psi q^{2}/\kappa ^{2}}
\mbox{ ,}
\end{equation}
and by Tarko and Fisher\cite{tf75} (TF) for $T<T_N$,
\begin{equation}
f(q/\kappa ) \propto \frac
{(1+\phi ^{\prime \,2}q^{2}/\kappa ^{2})^{\sigma+\eta/2}}
{(1+\psi ^\prime  q^{2}/\kappa ^{2})(1+\phi ^{\prime \prime\, 2}q^{2}/
\kappa ^{2})^{\sigma}}
\mbox{ ,}
\end{equation}
where $\phi$, $\phi ^{\prime}$, $\phi^{\prime \prime}$,
$\sigma$, $\psi = 1 + 1/2 \eta \phi ^2$ and
$\psi ^{\prime}=1+1/2 \eta {\phi ^{\prime}}^2 +\sigma ({\phi ^{\prime}}^2-
{\phi ^{\prime \prime}}^2)$ are fixed to values determined from
the numerical studies.  The values are given in Table I.  The expressions have
the correct scaling behavior in the limits $|q|/\kappa \rightarrow 0$ and
$|q|/\kappa \rightarrow \infty $ and serve as appropriate
interpolation functions between those limits.  We show the
critical exponents and amplitude ratios obtained from
fits of the data obtained for $10^{-4} < |t| < 10^{-2}$
using these scaling forms in Table I as well as theoretically determined
universal critical parameters.
The experimental and theoretical values
serve to contrast those from similar analyses done on the
diluted system $Fe_{0.93}Zn_{0.07}F_2$ to be discussed next.  The
corrections to MF are only significant for $|t|<10^{-3}$ and are most
significant for $T<T_N$.  Since many $d=3$
studies do not probe critical behavior any closer than this to $T_c$, the
Lorentzian line shape generally serves satisfactorily for
extracting estimates of the critical exponents.
Obviously, more precise measurements that probe regions of smaller $|t|$
can yield much more accurate critical parameters, but only if a suitable
line shape is used.

Data for the pure $d=2$ case of $K_2CoF_4$ has been analyzed in a similar
manner\cite{chb84}.  In this case $\eta = 0.25$ and an analysis using
the MF Lorentzian fails markedly below $T_N$.  The appropriate
TF and FB equations,
in contrast, yield critical behavior consistent with the $d=2$ Ising model.

The scattering function for the dilute antiferromagnet
is predicted to have an additional term not present in
the pure case\cite{pa85} that may affect experimentally determined
to corrections to scaling
and, possibly, measurements of the amplitude ratio for $\chi _s^{}$.
The exponents are probably not influenced since this extra contribution
vanishes as $T \rightarrow T_N$.  A fit to a simple Lorentzian
seems to work reasonably well for data with $|t|<10^{-3}$
in the very dilute antiferromagnet $Fe_{0.46}Zn_{0.54}F_2$.
For the random-exchange Ising model\cite{cprv99}, $\eta \approx 0.04$, which
is similar to the value for the pure case.  Hence, it is reasonable
that the Lorentzian line shape works well in this reduced temperature
range just as it does in the pure $d=3$ Ising case.  For data closer to
$T_c$, however, just as in the pure case, we would expect deviations
from the MF Lorentzian, particularly for $T<T_c$.  Unlike the
Tarko-Fisher case for the pure Ising model,
no approximants have been worked out for the random-exchange model
for use in data analyses beyond MF.  One possible approach is to
use the same TF and FB expressions developed for the pure case.
This assumes that the line shapes for the pure and REIM are
very similar.  Such an approach was successfully employed to analyze the
data for the dilute $d=2$ antiferromagnet\cite{hcni87}.
Another strategy is to use the same forms, but to let the TF/FB
parameters be free fitting parameters.  Such functions would satisfy
scaling in the proper limits and would hopefully be very good interpolative
functions between those limits.  We will describe below results for
$Fe_{0.93}Zn_{0.07}F_2$ obtained using the latter of these two methods.

We next turn to the more difficult case of the random-field
scattering in $Fe_{0.93}Zn_{0.07}F_2$, which occurs for
$H>0$.  In the case studied $H=7T$.  We expect the scattering
function for $|q|>0$ to be
\begin{equation}
S(q) = A^{\pm}\kappa ^{\eta-2}f(q/\kappa) +B^{\pm}{A^{\pm}}^2\kappa
^{\bar{\eta} -4}g(q/\kappa) \quad .
\end{equation}
Taking into consideration the instrumental
resolution as well as the two separate scaling functions, there
is little chance of using the data directly to determine
the two independent scaling functions.  Hence, we must start with
model functions and test their appropriateness.  The first natural
test functions
to use in the data analysis are the MF scaling ones\cite{gmm77},
\begin{equation}
S(q) = \frac{A^{\pm}\kappa ^{-2}}{1+q^2/\kappa ^2} +
\frac{B^{\pm}\kappa ^{-2}}{({1+q^2/\kappa ^2})^2} \quad ,
\end{equation}
where we might expect the amplitudes $A^{\pm}$ and $B^{\pm}$
to be temperature dependent, since $\eta$ and $\bar\eta$
are not, in fact, zero.  We did such an analysis
previously\cite{sbf98} and found reasonable fits
at all temperatures in the sense that the line shapes yielded
values of $\kappa$ and $\chi _s^{} = A^{\pm}\kappa ^{-2}$.  However,
when an attempt was made to fit these values to power law
behaviors, reasonable results were obtained above the transition
but not below.  The results above
$T_c(H)$ are $\nu = 0.90 \pm 0.01$ and $\gamma = 1.72 \pm 0.02$. These values
are consistent with results above $T_c(H)$, obtained\cite{bkj85}
for $x=0.6$ using a MF analysis.  At these lower concentrations,
equilibrium critical behavior is not obtained below $T_c(H)$, but the
data analysis was done well above $T_c(H)$ where equilibrium prevails.

We note that the failure of the MF analysis below $T_c(H)$ is
very similar to the situation observed\cite{chb84} in
the pure $d=2$ antiferromagnet $K_2CoF_4$.  The MF equations
fail in that case because $d=2$ is far from the upper critical dimension
$d_u=4$ and $\eta = 0.25$ in contrast with the pure
$d=3$ case where $\eta=0.04$.  The TF expression
developed for $d=2$ served nicely for the data analysis and
agreement was found with theory.  Interestingly, the FB scaling function
for $T>T_N$ is not very different from the MF one.
For the $d=3$ RFIM, the value of $\eta$ is predicted to be even larger,
perhaps as large as $\eta=0.5$.  Hence, it is not that surprising that
the MF data analysis fails below $T_c(H)$ in this case.  To proceed,
we must go beyond a simple MF line shape analysis.

Since there is not yet a theoretical line shape available beyond MF, we must
try to use scaling properties to guide us.  However,
if there are two independent scaling functions, as in Eq. 24,
the task becomes formidable.  Fortunately, there are two
approximations motivated by theoretical\cite{gaahs96}
and simulation\cite{bfmspr98} works.
The first is that $\bar\eta = 2 \eta$, a limiting case of the
Schwartz-Soffer\cite{ss86} inequality $\bar\eta \le 2 \eta$.
The second is that $g(q/\kappa)= f^2(q/\kappa)$ is a very good approximation.
We adopt these two simplifications, making the scattering function for $|q|>0$
the more manageable expression
\begin{equation}
S(q) = A^{\pm}\kappa^{\eta-2}f(q/\kappa)(1
+B^{\pm}A^{\pm}\kappa^{\eta-2}f(q/\kappa)) \quad .
\end{equation}
It is not clear at this point how accurate these approximations are.
However, it is not possible to proceed without them and they appear
to be well justified.  It is highly unlikely that experiments will
be able to test the validity of the assumptions directly. Only further theoretical progress can provide
a better starting point for the data analysis.
Note that the MF expression in Eq.\ 24 is a special case of this for $\eta=0$.
We are still faced with the correction
to instrumental resolution, however, which itself depends on the
line shape we are trying to determine\cite{by87}.  Hence, it is still hard
to scale the data directly without an explicit functional form for
$f(q/\kappa)$.  To simplify the procedure further,
therefore, we have adopted as our scaling functions the
TF and FB scaling functions except that we allow all the
parameters to vary.  Hence, we are assured of the proper scaling
in the limiting cases of $|q| \rightarrow 0 $ and $ \kappa \rightarrow 0$
and hopefully the interpolation between these limits will be adequate
with the parameters determined from the fits of the data.
Unfortunately, the usual technique of fitting each scan separately, where a scan
is made in $q$ at fixed $T$,
and then extracting the exponents from power law fits to the
resulting $\kappa$ and $\chi _s^{}$ cannot work well since there
are now so many free parameters, including the various exponents,
amplitudes and the TF/FB parameters.  We can, however, fit all the
data scans simultaneously, since the line shape parameters are all the
same for every temperature for $T<T_c$ and $T>T_c$ and the critical
exponents are the same for all $T$.
The new technique has the advantage over the more classic technique
in that the line shape does not need to be known beforehand.  In
the RFIM, the line shape is both unknown and far from the MF
prediction, so the classic method using the MF line shapes
failed to yield the critical
behavior parameters.  Our new procedure using the TF and FB line shapes
with variable parameters and fitting all the data simultaneously,
on the other hand, was successfully employed
for the $H>0$ random-field
Ising behaviors in $Fe_{0.93}Zn_{0.07}F_2$ and yielded both the
critical behavior parameters and the critical scattering line shapes.
Both the classic and new techniques worked well for the $H=0$ case 
in which the line shapes are nearly mean-field and much simpler
than the RFIM ones.  This demonstrates the reliability of the new
technique and we can apply it with some confidence to the
RFIM, where the more classical technique fails to yield results.

\section{Experimental and Fitting Details} 

Adding to the difficulty of implementing scaling in the
scattering line shape analysis is the necessity to
account for significant instrumental resolution corrections.
The instrumental resolution can be measured for a
particular spectrometer configuration by measuring the width
of the magnetic Bragg scattering peak at low temperatures.
The Bragg scans in the transverse, longitudinal and vertical
directions well below the transition temperature yield
the response to scattering from the Bragg peak.
Theoretically, the Bragg peak
is a delta function and in practice is much more narrow than
the instrumental resolution in good quality crystals.  Since the
scattering in these anisotropic crystals is well described within
the static approximation, the scans in the three directions yield the
widths along the three principal axes of the resolution
ellipsoid.  If we have a theoretical line shape, we can use
the measured Bragg scattering scans to numerically
integrate the line shape which can then be compared directly with
the scattering data.
Alternatively, other groups have used Gaussian (or with less accuracy
triangular) approximations to the Bragg scans and then
analytically integrated to obtain the resolution corrections.
In this study we exclusively use
the numerical integration technique.
For resolution curves measured
at uniform steps in $q$, we have for the intensity in transverse data scans,
\begin{equation}
I(q) \sim \frac {\sum S((q-q_0-a)^2+b^2+c^2)T_aL_bV_c}{\sum T_aL_bV_c} \quad ,
\end{equation}
where the sums are over $a$, $b$, and $c$ and $q_0$ accounts for imperfect
alignment in the transverse direction.  For a well aligned crystal, $q_0$ is
usually much smaller than the resolution width in the transverse direction.
Misalignments along the vertical and longitudinal directions are
generally inconsequential for a well aligned crystal since the resolutions
are much courser in these directions than in the transverse one.
$T_a$, $L_b$ and $V_c$ are the approximately Gaussian Bragg line
shapes measured at low $T$ using evenly spaced steps in $q$.  This
is the technique explained\cite{by87} in detail in the context of
the study of $FeF_2$.
If a crystal has a small mosaic, this can also be approximately taken into
account in the same manner.

To fit the resolution-convoluted line shape to the
scattering data, we use a nonlinear least squares
fitting routine to determine the parameters of the
theoretical line shape and critical exponents.
Note that with each iteration of the
fitting program, the line shape must be reintegrated over the
resolution ellipsoid since the amount of correction
from the resolution convolution depends on the line shape and, hence,
the fitted parameters.
Since the resolution correction in neutron scattering experiments
is substantial and line shape dependent, it is very difficult to
determine the line shape directly from the data.  Hence,
in the absence of a theoretical model, we must choose a trial function
that satisfies the correct scaling requirements and allows
suitable flexibility in the fits of the data.
Our strategy is to use the TF and FB line shapes described above.

For the experiments we used two samples of $Fe_{0.93}Zn_{0.07}F_2$.
One is the same large sample used in specific heat experiments\cite{sb98}.
It is somewhat irregular in shape and has a mass of 1.35 g.
The magnetic concentration gradient limited the range of data
unaffected by rounding to $|t|>1.15 \times 10 ^{-3}$. 
The second is a slice cut from the large sample with its faces
perpendicular to the magnetic concentration
gradient. It is approximately one tenth the mass of the original
sample.  The smaller sample was used to obtain data closer to the
transition with $|t|>1.14 \times 10^{-4}$.
The data from both samples were used simultaneously in the data
fits with different instrumental resolution corrections
appropriate to the two samples and the spectrometer configurations
used to make measurements on them.
Neutron scattering measurements were made
at the Oak Ridge National Laboratory
High Flux Isotope Reactor using a two-axis spectrometer
configuration.  We used the (0 0 2) reflection
of pyrolytic graphite (PG) at an energy of $14.7$ meV
to monochromate the beam.  We mainly employed two different
collimation configurations.  The lower resolution,
primarily used for the large sample,
is with 70 min of arc before
the monochromator, 20 before the sample and
20 after the sample.  Primarily for the thin sample,
we made scans with 10 min of arc before and after the sample.
Two PG filters were used to eliminate
higher-order scattering.  The carbon thermometry scale was calibrated
to agree with recent specific heat results \cite{sb98} for the
$H=0$ transition.  The field dependence of the thermometry was also
calibrated.  All scans used in this report, other than those used for
obtaining the resolution ellipsoid, are transverse ones about
the (1~0~0) antiferromagnetic Bragg point.

\section{Fitting Results}

We applied the techniques described above to the random-exchange behavior ($H=0$),
where there is only one term in the scattering function for $|q|>0$.
We first did the fits using the MF line shapes imposed by setting
$\sigma = 1$ and $\phi$, $\phi '$ and $\phi ''$ equal to zero.
The results for the fitted critical parameters are essentially
identical to the results from the conventional method of fitting
each scan to obtain the temperature dependent correlation and susceptibility
and subsequently fitting them to extract the exponents and amplitude ratios.
The results are shown in Table I.  We include the
expressions for the normalized $\overline{\chi^2}$ in the table, which are most
useful for relative comparisons between fits.  The values are not
close enough to unity to consider the fitting functions statistically perfect
ones.  Very small systematic errors from the resolution corrections and
the approximations in the line shapes
can easily account for values of $\overline{\chi^2}$ being somewhat larger than
unity.  The goodness of the fits are better judged from the scaling
plots discussed below.  The agreement between the two
methods of fitting the data gives us some confidence in fitting
the field data, for which the line shape is unknown, using the
technique in which all data are simultaneously fit.  We proceed beyond
MF by doing the fits with the TF and BF parameters as free parameters.
The fitting results are again shown in Table II.  The range of data was
restricted to the reduced temperature range $|t|< 0.15$ and includes
2198 data.  The results for the
critical exponents and amplitudes are not very different from
the previous result and the line shape is not very different from the MF one.
When examining individual scans, we observe no systematic deviations of the data
from the fit at any temperature for which data were included in the fit.  We
show representative scans in Fig.\ 1, including data above and below
$T_c$.  No systematic deviations of the data from the fits are evident,
indicating that the line shapes work well for the data within the range
of $|t|$ used in the fits.  To demonstrate that the data are well
described by the fitted scaling
function, we subtract the scattering background from them, deconvolute them
with the instrumental resolution, divide them by $A^{\pm}\kappa ^{\eta -2}$
and plot them versus $|q|/\kappa$ in Fig.\ 2.  The data from the two samples
are plotted separately, but the solid curves are identical in the two cases.
The scatter of the data for the small sample simply reflects the smaller size
and the resulting lower count rate.  However, some of the data from the small
sample are taken much closer to $T_c$, since the rounding from the
concentration gradient is less significant.  The data close to $T_c$ are very
important in the fits.
The consistent results obtained using MF and scaling line shapes in the
analyses give us confidence that the scaling technique may work well even in
cases where the line shapes are very far from being MF, as in the RFIM case we
describe next.

For the RFIM case, the procedure is identical to the random-exchange one
above, except that the scattering function involves the more complicated
expression in Eq.\ 25.  In one fit, we restricted the data range to
$|t|<10^{-2}$, with 2444 data, since this is the temperature range over which
the asymptotic logarithmic specific heat behavior is observed\cite{sb98}.
In a second fit, we further restricted the temperature range to
$|t|<3 \times 10^{-3}$, with 1000 data, to test whether crossover effects were
still significant.  The fitting results are shown in Table III.  The two
fits are rather consistent, suggesting that the results are close to the
asymptotic ones.  In Fig.\ 3 we show data and the corresponding fits
for a few scans to show that there are no significant systematic deviations.
After determining the line shape
parameters and critical exponents, we used a procedure similar to that
described above to demonstrate that the data collapse onto scaling
functions for $T<T_c$ and $T>T_c$.  We subtracted the scattering background,
deconvoluted the data with the instrumental resolutions, divided by
$A^{\pm}\kappa ^{\eta -2}(1+B^{\pm}A^{\pm}\kappa^{\eta-2}f(q/\kappa))$
and, for clarity, plotted the results versus ${\kappa _o}^{\pm}|q|/\kappa$
instead of simply $|q|/\kappa$ so that the data for $T>T_c(H)$ and $T<T_c(H)$
do not overlap.  The results of this procedure are shown in Fig.\ 4.
The data from the large and small samples are shown separately.

We have obtained from the scaling analysis not only the critical parameters
but adequate approximations to the line shape scaling functions for both the
$d=3$ random-exchange and random-field Ising models.  We already have
the $d=2$ and $d=3$ pure line shapes from the TF and FB expressions.  It is
instructive to compare all of these to the MF Lorentzian line shapes.  The
comparisons are shown in Fig.\ 5, where we have plotted scaling functions
versus $|q|/\kappa$.  The upper plot is for $T>T_c$ and the lower one for
$T>T_c$.  The lowest curve in each case represents the simple Lorentzian line
shape that is accurate in the MF limit.  The pure $d=3$ Ising model line
shapes show nearly MF behavior for $T>T_c$.  For $T<T_c$, more significant
deviations from MF are apparent.  This is consistent with the experimental
results obtained using $FeF_2$.  The $d=3$ experimental random-exchange
scaling functions indicate that there is not much difference between the pure
and random-exchange line shapes for $d=3$.  In contrast, for
the random-field case, the deviations from MF found in the experiments are
very large for $T<T_c(H)$ in comparison to the pure and random-exchange $d=3$
Ising models.  Even in the case for $T>T_c(H)$ the deviations are large
relative to the other $d=3$ models.  The very large deviations for $T<T_c(H)$
are consistent with the failure of the MF Lorentzian line shape in producing
values for $\kappa$ and $\chi _s^{}$ that obey power law behavior.  For the pure
$d=2$ Ising model, shown as a reference for non-MF behavior, the
results for $T<T_c(H)$ are the largest of the examples shown, whereas the
behavior for $T>T_c$ is not far from MF.  We see from this comparison of the
scaling functions, particularly for $T<T_c$, that MF behavior is fairly well
followed for the $d=3$ pure and random-exchange models, whereas for the
random-field case, where the upper critical dimension has significantly
increased to $d_u=6$, the line shapes are very non-Lorentzian.  Determining
the line shape from the scattering data is difficult.  We have achieved
an approximate determination of the proper line shapes.  Once theoretical
results give a more firm foundation for the trial function for the scattering
line shape, we will be able to give a more concrete comparison of the
experiments to theory and simulations.

In Table IV we show the results of the critical exponents for the specific
heat and neutron scattering RFIM experiments compared with simulation results.
For the pure $d=3$ Ising model, the experimentally determined universal
exponents and amplitude ratios agree very well with theoretical and
simulation results.  The experimental results from neutron scattering
were obtained using the TF and FB line shapes.  The specific heat was obtained
using both pulsed heat and birefringence techniques and clearly
can be considered to be exceedingly well characterized.  
For the REIM, we again have excellent agreement between experiment and
Monte Carlo simulations.
The specific heat is determined most precisely using the birefringence
technique, but the results are consistent with the pulsed heat data.
This case can also be considered well characterized.

The RFIM case shows mixed results when the experimental exponents are
compared with those from simulations.  The scattering results for $\nu$
and $\gamma$ are quite consistent with the simulation values.  The
exponent $\beta$ has not been reliably measured yet and its comparison
with simulation results is very important.  The most glaring
inconsistency is between the simulation and experimental values
of $\alpha$, where the experiments indicate a symmetric, logarithmic
divergence and the simulations indicate a non-divergent peak.
This inconsistency deserves further study.

Although we have achieved fits to the data obtained in equilibrium scattering
experiments using $Fe_{0.93}Zn_{0.07}F_2$, the asymptotic behavior is only
observed very close to $T_c(H)$ at $H=7$~T.  Our results may be somewhat
influenced by the effects of crossover to random-exchange behavior.  We hope
in the future to do these experiments at much higher fields to ensure that the
results we have obtained in this study are close to the asymptotic ones.  In
addition, by using apparatus capable of reaching much higher fields, we can
study the scaling behavior of the scattering line shapes as a function of    
$|t|H^{-2/\phi _{RF}}$ in a similar way to our treatment of the specific heat.

This work has been supported
by DOE Grant No. DE-FG03-87ER45324
and by the Oak Ridge National Laboratory,
which is managed by UT-Battelle, LLC, for the
U.S. Dept. of Energy under contract DE-AC05-00OR22725.

\begin{figure}
\centerline{\hbox{
\psfig{figure=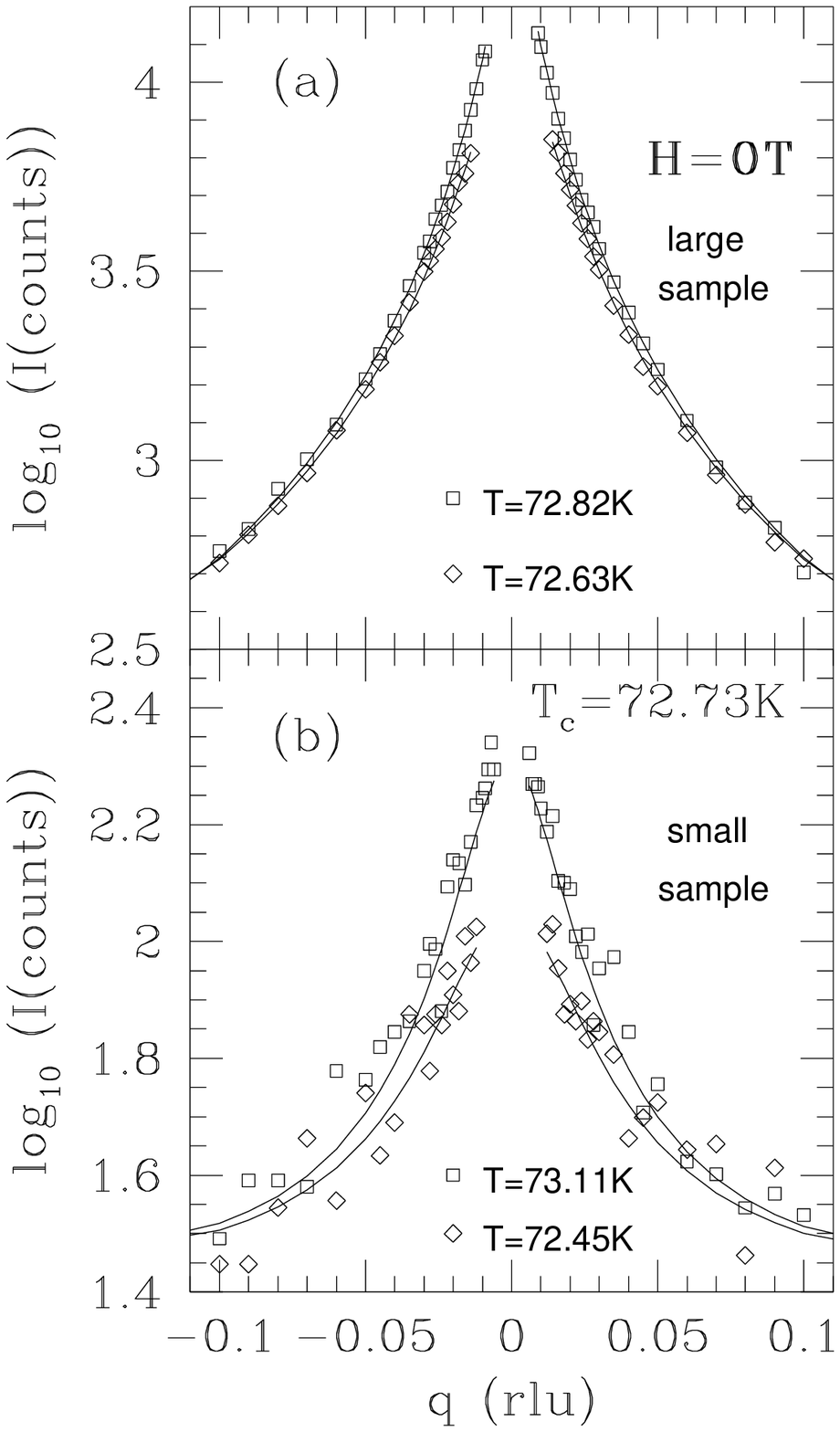,height=6.0in}
}}
\caption{
The logarithm of the scattering intensity vs.\ $q$ and the curves
representing the corresponding fits for $H=0$.  One scan for $T>T_c(H)$ and
one for $T<T_c(H)$ are shown for the large sample in figure a and
similarly for the small sample in figure b.}
\end{figure}

\begin{figure}
\centerline{\hbox{
\psfig{figure=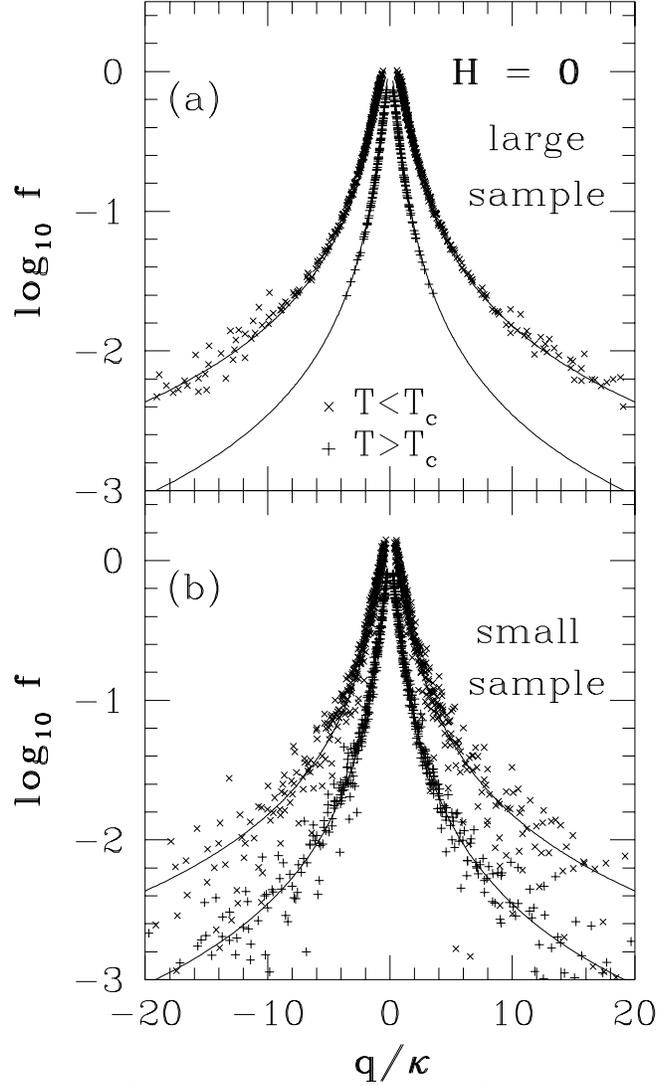,height=6.0in}
}} 
\caption{
Scaled neutron scattering data, deconvoluted with the instrumental resolution,
taken at different temperatures at
$H=0$ T collapsed onto the universal function $f(q/\kappa )$.
The scatter in the small sample data is larger
due to smaller number of counts obtained in the thin sample. The
fit was made for $|t|<0.01$.  The solid curves, which are identical
in the upper and lower plots, represent the line shapes for
$T>T_c$ and $T<T_c$ determined from the fits to the data.}
\end{figure}

\begin{figure}
\centerline{\hbox{
\psfig{figure=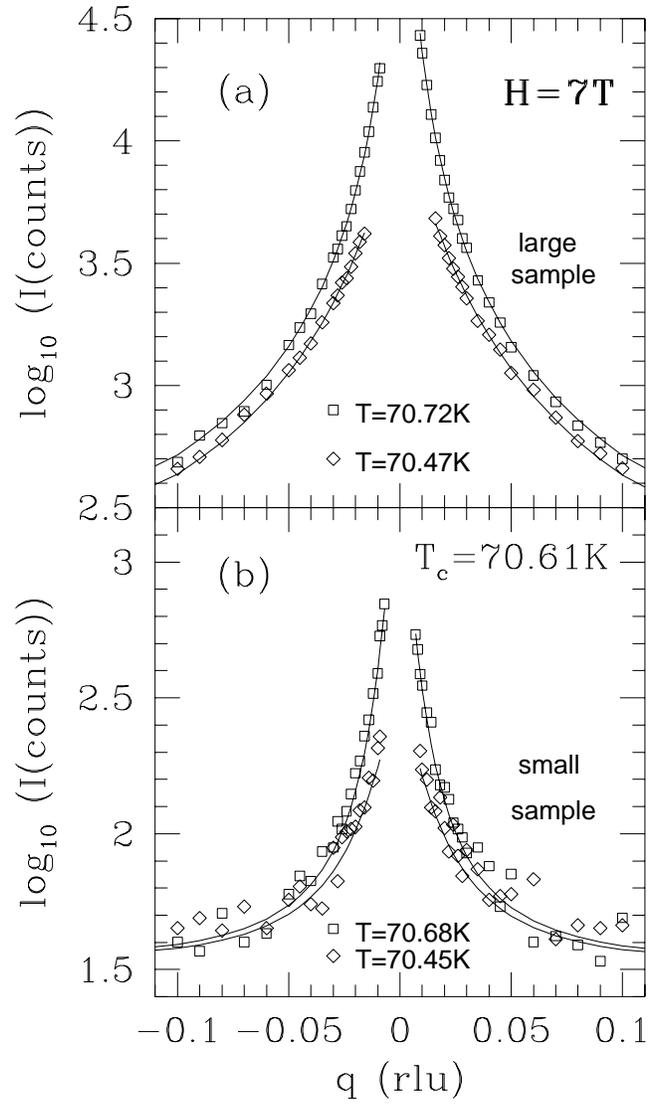,height=6.0in}
}} 
\caption{
The same as in Fig.\ 1, but for the the random-field Ising case $H=7$~T.}
\end{figure}

\begin{figure}
\centerline{\hbox{
\psfig{figure=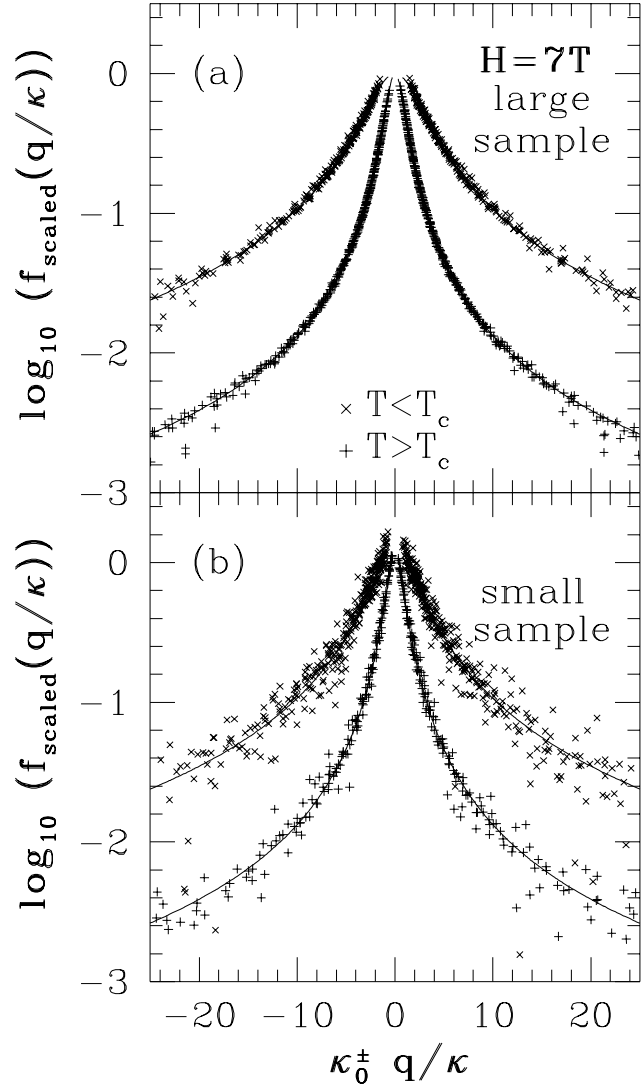,height=6.0in}
}} 
\caption{
The same as in Fig.\ 2, but for the random-field Ising case $H=7$~T.
The fit was made for $|t|<0.01$.}
\end{figure}

\begin{figure}
\centerline{\hbox{
\psfig{figure=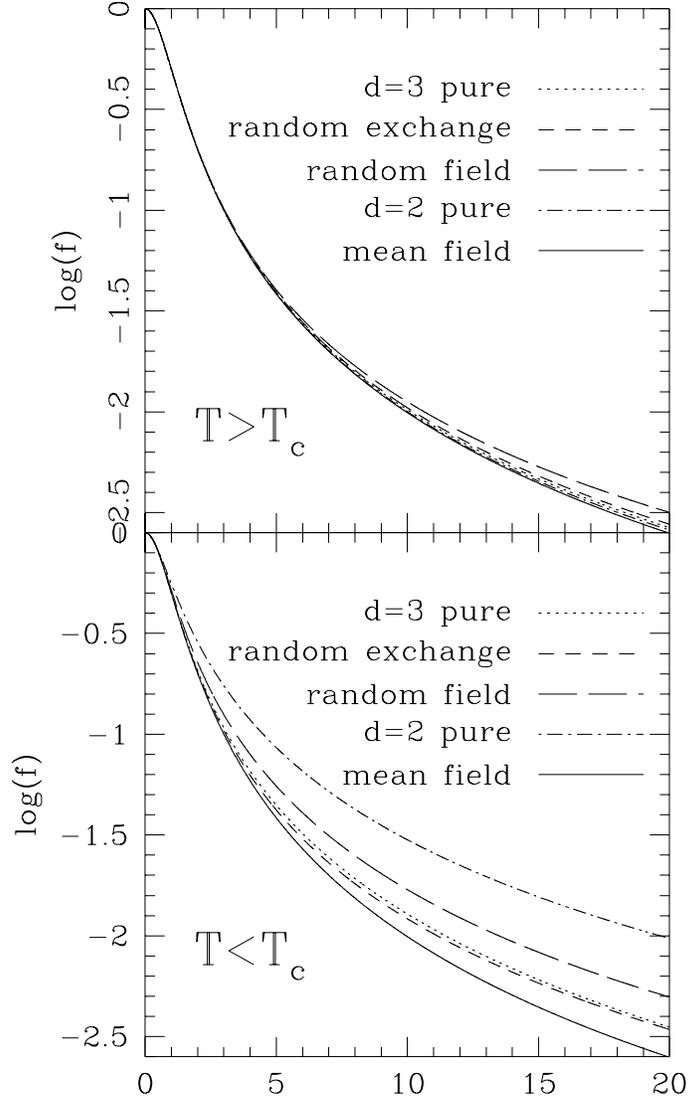,height=6.0in}
}} 
\caption{
A comparison of the logarithm of the scaling functions
versus $q/\kappa$ for different models.
The pure cases are from the TF and FB expressions.
The REIM and RFIM are from the experiments, as described in
the text.  Note that the corrections to the MF
equation are largest below the transition.  The random-field
deviations for $d=3$ are greater than the pure
and random-exchange, but are significantly smaller
than the pure $d=2$ scaling function below the transition.
Above the transition, all of the line shapes are close to MF.}
\end{figure}

\newpage

\begin{table}
\begin{center}
\begin{tabular}{lcc}
\hline
parameter       & FB and TF & $d=3$ Ising\cite{cprv99} \\
\hline
$T_c$                    &   $78.418 \pm 0.001$~K&      \\
$\nu$                   &   $0.64 \pm 0.01$       &   $0.63002 \pm 00023$  \\
$\kappa^+_0/\kappa^-_0 $ &   $0.53 \pm 0.01$       &   $0.510 \pm 0.002$   \\
$\gamma$                &   $1.25 \pm 0.02$     & $1.2371 \pm 0.004$ \\
$\chi^+_0/\chi^-_0$     &   $4.6 \pm 0.2$       & $4.77 \pm 0.02$ \\
$\eta$			&   $0.056$*  \\
$\psi$ & $1.001$*  \\
$\sigma$                &   $2\eta = 0.111$*   \\
$\phi$                  &   $0.15$*  \\
$\phi^\prime$           &   $0.3247$*   \\
$\phi^{\prime \prime}$  &   $0.09355$*   \\
$\psi^{\prime} $		&   $1.0137$*
\end{tabular}
\end{center}
\caption[Fitting Parameters for the $H=0$ T Case]
{
Experimental and Theoretical values for the critical parameters
of the $d=3$ pure Ising model.  Data were fit over the range
$10^{-4}<|t|<10^{-2}$.  Note that $\psi = 1 + 1/2\eta \phi ^2$
and $\psi^{\prime} = 1 + 1/2\eta \phi ^{\prime\,2} + \sigma
(\phi^{\prime\,2}-\phi^{\prime \prime, 2})$.  Values marked with
$^{*}$ are fixed parameters in the fitting process.
}
\end{table}

\begin{table}[ht!]
\begin{center}
\begin{tabular}{lcc}
parameter       &       Lorentzian   &  TF/FB \\
\hline

$T_c$ (fixed)                   &   $72.73K^*$          &   $72.73K^*$      \\
$\eta$                  &   $0.079\pm0.010$     & $0.079\pm0.012$ \\
$\nu$                   &   $0.70\pm0.03$       & $0.70\pm0.02$  \\
$A^+$                   &   $7.66\pm0.15$       & $7.59\pm0.10$   \\
$A^-$                   &   $6.71\pm0.10$       & $6.42\pm0.10$   \\
$\kappa^+_0$            &   $0.57\pm0.02$       & $0.56\pm0.02$   \\
$\kappa^-_0$            &   $1.21\pm0.10$       & $1.13\pm0.05$   \\
$\sigma$                &   $1^*$               & $0.16\pm0.20$   \\
$\phi$                  &   $0^*$               & $0.18\pm0.02$   \\
$\phi^\prime$           &   $0^*$               & $0.18\pm0.02$   \\
$\phi^{\prime \prime}$  &   $0^*$               & $0.08\pm0.10$   \\
$C$			&   $0.0142\pm 0.0002$	& $0.0142\pm0.0001$\\
$\overline{\chi^2}$     &   $2.0$               & $1.7$           \\
No. pts.              &       $2280$            & $2198$                  \\
\end{tabular}
\end{center}
\caption[Fitting Parameters for the $H=0$ T Case]
{
The values found for the parameters from the fits for the $H=0$ T data.
The exponents and amplitude ratios are defined in the text.
$C$ refers to a constant, $q$-independent background scattering term.
The scattering data for the large sample were fit for
$|t| > 1.15\times10^{-3}$ and
the data for the small sample were fit for $|t|> 1.14\times10^{-4}$.
$T_c$ was fixed in the case of the scattering fits.
The first column of results is for the MF line shapes.  The second
is TF/FB using the pure line shape parameters.  The third column is
obtained by letting the TF/FB parameters be fit along with the exponents.
}
\end{table}

\begin{table}[ht!]
\begin{center}
\begin{tabular}{lcc}
parameter       &       $|t|<10^{-2}$   & $|t|<3\times10^{-3}$ \\
\hline

$T_c$ (fixed)                   &   $70.61K^*$          &   $70.61K^*$      \\
$\eta$                  &   $0.20\pm0.05$     &   $0.16\pm0.06$ \\
$\nu$                   &   $0.88\pm0.05$       &   $0.87\pm0.07$  \\
$A^+$                   &   $10.0\pm0.2$       &   $9.21\pm0.3$   \\
$A^-$                   &   $6.15\pm0.14$       &   $4.45\pm0.15$   \\
$\kappa^+_0$            &   $1.13\pm0.04$       &   $0.95\pm0.17$   \\
$\kappa^-_0$            &   $3.24\pm0.11$       &   $2.78\pm0.5$   \\
$B^+$                               &       $(4.7\pm0.1)\times10^{-5}$ &
$(3.00\pm0.13) \times10^{-5}$           \\
$B^-$                               &
$(4.0\pm0.3)\times10^{-5}$ & $(8.0\pm1.0) \times 10^{-5}$            \\
$\sigma$                &   $0.67\pm0.5$               &   $0.86\pm0.6$   \\
$\phi$                  &   $0.16\pm0.04$               &   $0.08\pm0.01$   \\
$\phi^\prime$           &   $0.39\pm0.25$               &   $0.36\pm0.3$   \\
$\phi^{\prime \prime}$  &   $0.31\pm0.25$               &   $0.26\pm0.2$   \\
$C$		&   $0.017\pm 0.001$		&   $0.016\pm 0.001$\\
$\overline{\chi^2}$     &   $3.07$               &   $2.3$           \\
No. pts.              &       $2444$                          &
$1000$                  \\
\end{tabular}
\end{center}
\caption[Fitting Parameters for the $H=7$ T Case]
{
The values found for the parameters from the TF/FB fits for the $H=7$ T data.
The data for the large sample were fit for $|t| > 1.15\times10^{-3}$ and
the data for the small sample were fit for $|t|> 1.14\times10^{-4}$.
$T_c$ was fixed in the case of the scattering data.
}
\end{table}

\begin{table}[ht!]
\begin{center}
\begin{tabular}{lcc}
Pure & $FeF_2$ & Renormalization and\\
&&High T Expansions\\
$\alpha$    & $0.11 \pm 0.005$\cite{bnkjlb83} & $0.1099 \pm 0.0007$\cite{cprv99}\\
&&$0.109 \pm 0.004$\cite{gz98}\\
$\beta$ & $0.325 \pm 0.005$\cite{wb67} & $0.32648 \pm 0.00018$\cite{cprv99}\\
&&$0.3258 \pm 0.0014$\cite{gz98}\\
$\nu$ & $0.64 \pm 0.01$\cite{by87} & $0.63002 \pm 0.00023$\cite{cprv99}\\
&&$0.6304 \pm 0.0013$\cite{gz98}\\
$\gamma $ & $1.25 \pm 0.02$\cite{by87} & $1.2371 \pm 0.0004$\cite{cprv99}\\
&&$1.2396 \pm 0.0013$\cite{gz98}\\
$\eta $ & $0.05 $\cite{by87} & $0.0364 \pm 0.0004$\cite{cprv99}\\
&&$0.0335 \pm 0.0025$\cite{gz98}\\
\hline \hline

Random & $Fe_xZn_{1-x}F_2 $ & Monte Carlo \\
Exchange & ($H=0$)  &  \\
$\alpha$    & $-0.10 \pm 0.02$\cite{sb98}& $-0.051 \pm 0.013$\cite{bfmspr98}\\
$\beta$ & $0.350 \pm 0.009$\cite{rklhe88}& $0.3546 \pm 0.0028$\cite{bfmspr98}\\
$\nu$ & $0.69 \pm 0.01$\cite{bkj86}& $0.6837 \pm 0.0053$\cite{bfmspr98}\\
$\gamma $ & $1.31 \pm 0.03$\cite{bkj86}  & $1.342 \pm 0.010$\cite{bfmspr98}\\
$\eta $ & $0.10$\cite{bkj86} & $0.0374 \pm 0.0045$\cite{bfmspr98}\\
\hline \hline

Random & $Fe_xZn_{1-x}F_2 $ & Monte Carlo \\
Field & ($H>0$) &  \& Exact Ground State\\
$\alpha$    & $0.0 \pm 0.02$\cite{sb98}& $ -0.5 \pm 0.2$\cite{r95}\\
&&$-0.55 \pm 0.2$\cite{nue97}\\
$\beta $ & not measured\cite{rkjr88} & $0.00 \pm 0.05$\cite{r95}\\
&&$0.02 \pm 0.01$\cite{nue97}\\
$\nu$ & $0.88 \pm 0.05$ & $1.1 \pm 0.2$\cite{r95}\\
&&$1.14 \pm 0.10$\cite{nue97}\\
$\gamma $ & $1.58 \pm 0.13$ & $1.7 \pm 0.2$\cite{r95}\\
&&$1.5 \pm 0.2$\cite{nue97}\\
$\eta $ & $0.20 \pm 0.05$ & $0.50 \pm 0.05$\cite{r95}\\
$\bar{\gamma}$ & $2\gamma = 3.16 \pm 0.26$ & $3.3 \pm 0.6$\cite{r95}\\
&&$3.4 \pm 0.4$\cite{nue97}\\
$\bar{\eta}$ & $2\eta = 0.40 \pm 0.10$ & $1.03 \pm 0.05$\cite{r95}\\
\hline
 \hline
\end{tabular}
\end{center}
\caption[Critical exponents for the $d=3$ pure, REIM and RFIM]
{
The $d=3$ Ising critical exponents obtained from experiments on
the $Fe_xZn{1-x}F_2$ and the corresponding values from Monte Carlo
simulations and exact ground state calculations.  More extensive
experimental, theoretical and simulation results are compared
for the pure model in ref.\ \cite{gz98} and \cite{cprv99} and
for the random-exchange model in ref. \cite{bfmspr98} and \cite{fhy99}.  
Note that, unlike
the pure and REIM cases, it is well established that hyperscaling
is violated in the RFIM case\cite{by91}, i.e. $\alpha + d\nu \ne 2$
for the RFIM.}

\end{table}


\begin{references}
\bibitem{rkjr88}The only measurement is using the dilation
technique at lower concentration were the system is not
in equilibrium by C. A. Ramos, A. R. King, V. Jaccarion and S. M. Rezende,
J. de Phys. {\bf 49}, C8-1241 (1988).
\end{references}
\end{document}